\documentclass[runningheads]{llncs}
\usepackage{graphicx}
\begin{document}
\title{Semantic Segmentation of Neuronal Bodies in Fluorescence Microscopy Using a 2D+3D CNN Training Strategy with Sparsely Annotated Data}
\titlerunning{Semantic Segmentation of Neuronal Bodies Using a 2D+3D Strategy}
%
\author{Filippo M. Castelli\inst{1}\and
Matteo Roffilli\inst{3}\and
Giacomo Mazzamuto\inst{1,2}\and\\
Irene Costantini \inst{1,2}\and
Ludovico Silvestri \inst{1}\and
Francesco S. Pavone \inst {1,2,4}}
\authorrunning{F. M. Castelli et al.}
%
\institute{
European Laboratory for Non-Linear Spectroscopy (LENS), University of Florence\\Via Nello Carrara 1 50019 Sesto Fiorentino (FI), Italy\\\email{castelli@lens.unifi.it}\and
National Institute of Optics, National Research Council (INO-CNR)\\Via Nello Carrara 1, 50019 Sesto Fiorentino (FI), Italy\and
Bioretics SrL, Corte Zavattini, 11, 47522 Cesena (FC), Italy \and
Department of Physics and Astronomy, University of Florence\\Via G. Sansone, 1 50019 Sesto Fiorentino, Italy}
\maketitle              
\begin{abstract}
Semantic segmentation of neuronal structures in 3D high-resolution fluorescence microscopy imaging of the human brain cortex can take advantage of bidimensional CNNs, which yield good results in neuron localization but lead to inaccurate surface reconstruction. 3D CNNs on the other hand would require manually annotated volumetric data on a large scale, and hence considerable human effort.
Semi-supervised alternative strategies which make use only of sparse annotations suffer from longer training times and achieved models tend to have increased capacity compared to 2D CNNs, needing more ground truth data to attain similar results.
To overcome these issues we propose a two-phase strategy for training native 3D CNN models on sparse 2D annotations where missing labels are inferred by a 2D CNN model and combined with manual annotations in a weighted manner during loss calculation.

\keywords{semantic segmentation \and neuronal segmentation \and pseudo-labeling.}
\end{abstract}
\section{Introduction and Related Work}
Quantitative studies on the human brain require the ability to reliably resolve neuronal structures at a cellular level. Extensive three-dimensional imaging of the cortical tissue is nowadays possible using high-resolution fluorescence microscopy techniques which are characterized by high volumetric throughput and produce Terabyte-sized datasets in the form of 3D stacks of images.

Semantic segmentation of neuronal bodies in such datasets can be tackled with two-dimensional CNNs: this yields good results in neuron localization, but leads to inaccurate surface reconstruction when the 2D predictions are combined, in a z-stack approach, to estimate a 3D probabilistic map.
The issue could be solved by relying on 3D models such as 3D U-Net by Cicek et al. \cite{cicek_3dunet} which in turn requires large-scale annotation of volumetric ground truth and consequently an unfeasible human efforts. Cicek's own semi-supervised solution introduces a voxelwise binary loss weightmap that excludes the contributions of non-labeled voxels. However, the restriction of loss contributions to a small fraction of the training data is highly inefficient and leads to long training times.
To overcome these issues, namely the overwhelming effort for manual annotations and the optimization inefficiency due to very sparse input data, we propose strategy for training 3D CNN models by following \textit{pseudo-labeling} framework introduced by Lee et al. \cite{lee_pseudolabel}. In our setup available partial annotations (ground truth) are complemented on unlabeled areas bypseudo-labels generated by an independent 2D CNN model trained on available annotations.
We are exploring the application of this training scheme in semantic segmentation of high-resolution Two Photon Fluorescence Microscopy imaging of the human cortex \cite{costantini_biorxiv}\cite{mazzamuto_automatic}.

\section{Methodology}
Three dimensional imaging of four different $mm^3$-sized human neocortical tissue samples was acquired  using a custom-made Two Photon Fluorescence Microscope, as described in \cite{costantini_biorxiv}. The images were acquired in the form of multiple, partially overlapping $512\times512 px$ stacks with a voxel size of $0.88\times0.88\times2 \mu m^3$ which are  fused using a custom stitching tool. Four volumes sized $100\times100\times100 \mu m^3$ ($114\times114\times50 px$) were manually annotated by an expert operator in a 2D slice-by-slice fashion, three samples were chosen for model training and the fourth was kept for testing.
The voxel content of our volumetric dataset can be split into a \textit{labeled} and an \textit{unlabeled} subset as $X = X_L + X_U$: the goal of our semi-supervised learning task is to jointly use the set of all labeled datapoints $\{X_L, Y_L\}$ and the set of unlabeled ones $X_U$ to train a segmentation model for inductive inference on previously unseen data.
\begin{figure}[!hbt]
\includegraphics[width=\textwidth]{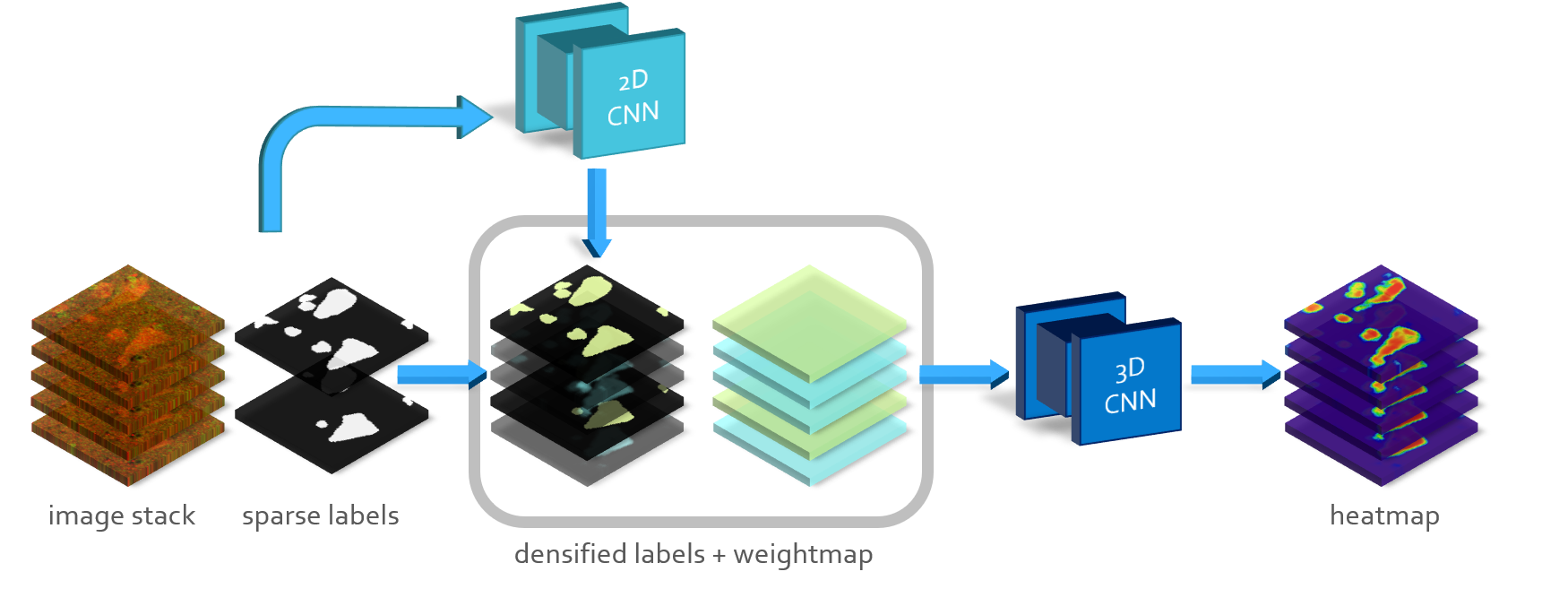}
\caption{Proposed training scheme: a 2D CNN model is used to infer pseudo-labels in the unlabeled parts of the training dataset, the main 3D CNN model is trained on a dense weighted combination of ground truth labels and pseudo-labels.} \label{fig_flow}
\end{figure}
Our strategy makes use of a 2D CNN segmentation model trained on the available ground truth, this model is used to transductively infer pseudo-labels $\hat{Y}_U$ on the unlabeled dataset partition. Another 3D CNN model $f_{\theta}$ is then trained on the dense combination $\{X_L, Y_L, X_U, \hat{Y}_U \}$ of available ground truth and inferred pseudo-labels using a weighted loss function
\begin{equation}\label{eq:loss}
L(X_L, Y_L, X_U, \hat{Y}_U; \theta) =  L_s (X_L, Y_L; \theta) + \alpha L_{p}(X_U, \hat{Y}_U; \theta) 
\end{equation}
where loss terms $L_s$ and $L_p$ are respectively a \textit{supervised} term that applies only to the labeled partition of the dataset and a \textit{pseudo-label} term that only applies to the \textit{unlabeled} one, while $\alpha$ is a tradeoff parameter. We chose to use for both terms a \textit{binary cross-entropy} loss function: this allows for straightforward implementation using a voxel-wise loss weightmap with values $1$ for labeled voxels and $\alpha$ for unlabeled ones.\\
We test our pseudo-labeling scheme on a 3D CNN model based on 3D UNET, trained on $64\times64\times32$ patches. The auxiliary pseudo-labeling CNN is a lightweight 2D CNN with three convolutional layers, followed by two fully connected layers which we have already used in segmentation of large Two Photon Microscopy imaging datasets in Costantini et al. \cite{costantini_biorxiv}.

\section{Evaluation and Results}
\begin{figure}[!hbt]
    \centering
    \includegraphics[width=\textwidth]{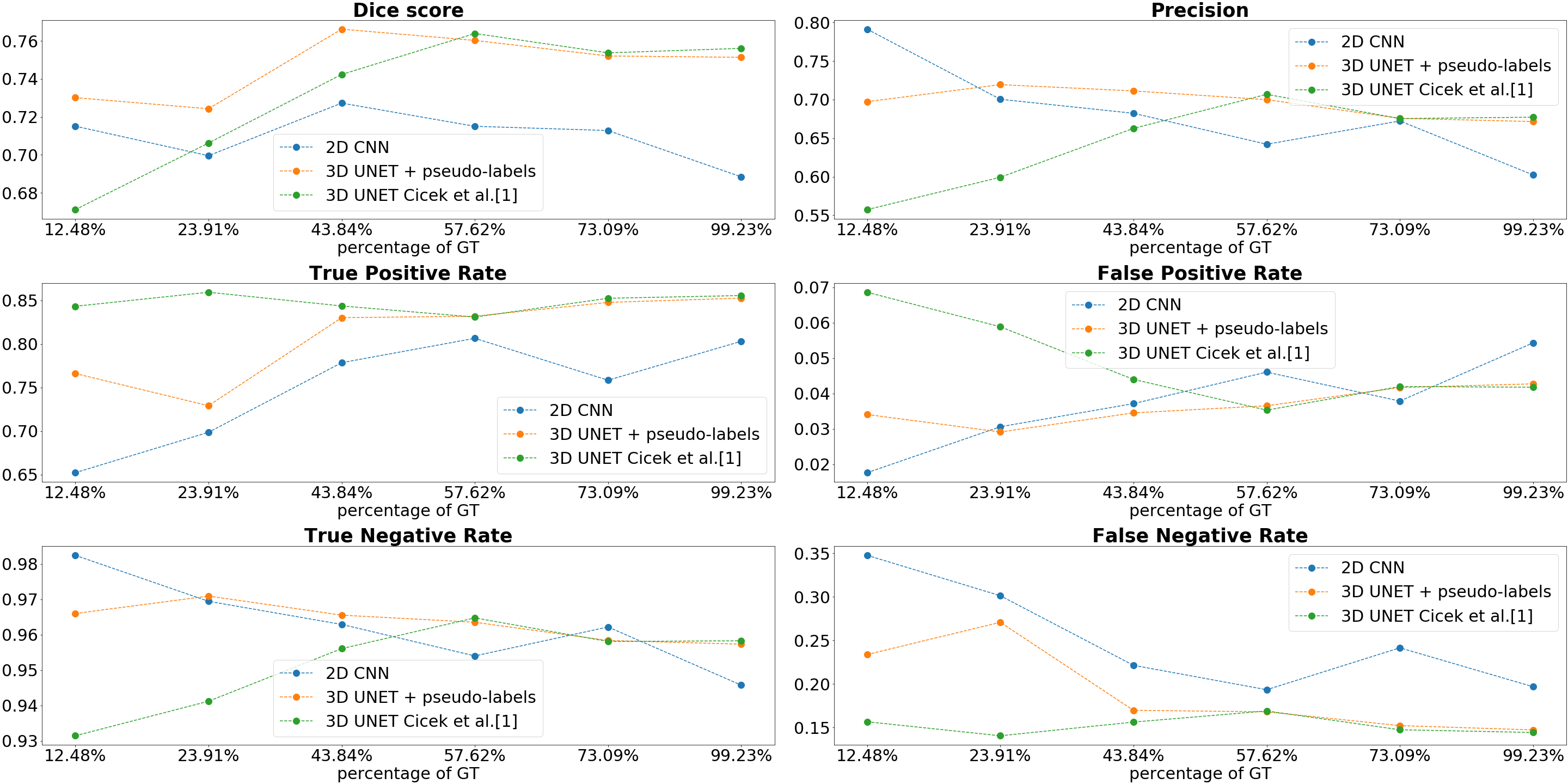}
    \caption{Performance on an independent test set at various label sparsity conditions}
    \label{fig:metrics}
\end{figure}

The effectiveness of our semi-supervised approach was tested by simulating several conditions of label sparsity by using increasing numbers of 2D slices as ground truth, while treating the remaining voxels as unlabeled. We compare the performances of the pseudo-labeling model, the 3D model trained with the described pseudo-labeling scheme and the same network, sparsely trained without pseudo-labels on the available annotations.
By choosing a binary classification threshold of $0.5$, in Figure \ref{fig:metrics} we report on a voxel basis the statistical assessment of the semantic segmentation quality as Dice score, as well as the values of precision and the confusion matrix evaluated on the test dataset.

\section{Conclusions}
The results suggest that in heavily sparse labeling conditions, the model trained with our semi-supervised scheme could achieve better segmentation performances than both the 2D CNN model and the same 3D CNN model trained without pseudo-labels.
\begin{figure}[!hbt]
    \centering
    \includegraphics[width=\textwidth]{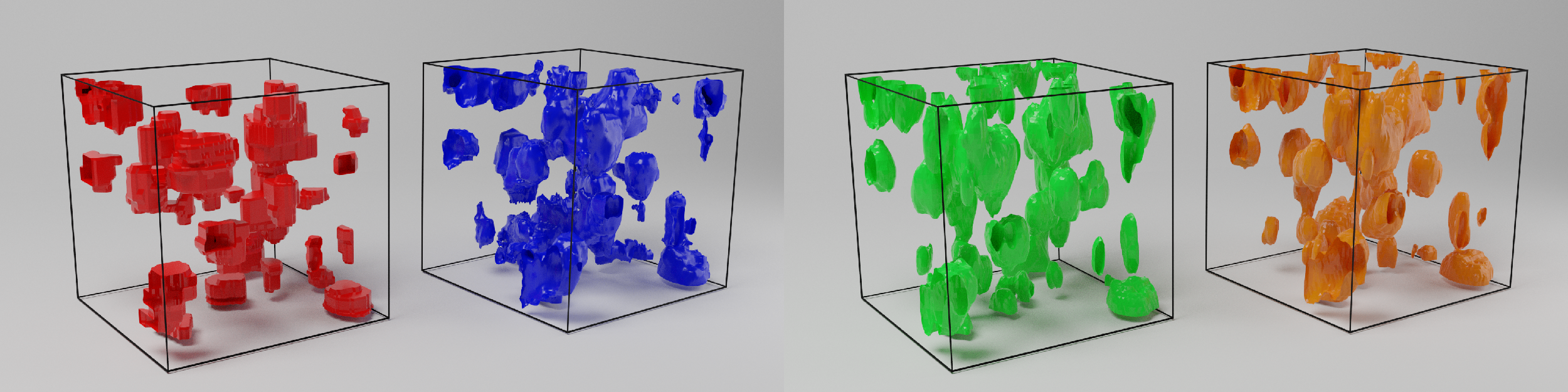}
    \caption{Visual comparison between training schemes at $43.84\%$ of annotated voxels, from left to right: GT annotations + 3D isosurface reconstruction, 2D CNN model + 3D isosurface reconstruction, 3D CNN sparse training + 3D isosurface reconstruction, 3D CNN sparse training with pseudo-labels + 3D isosurface reconstruction.}
    \label{fig:visual_reference}
\end{figure}
Visually, as depicted in Figure \ref{fig:visual_reference}, the 3D CNN model trained with our scheme achieves better reconstruction of the neurons' marginal sections and yield more consistent results when compared to both the pseudo-labeling model and the baseline 3D training without pseudo-labels.\\
\textbf{Acknowledgements:} Research reported in this work was supported by The General Hospital Corporation Center of the National Institutes of Health under award number 1U01MH117023-01. The content is solely the responsibility of the authors and does not necessarily represent the official views of the National Institutes of Health.

%
%
%

\begin{thebibliography}{4}

\bibitem{cicek_3dunet}
Çiçek, Ö. et al.: 3D U-Net: Learning Dense Volumetric Segmentation from Sparse Annotation. In: Medical Image Computing and Computer-Assisted Intervention - MICCAI 2016. (2016)

\bibitem{costantini_biorxiv}
Costantini, I. et al.: A combined pipeline for quantitative analysis of human brain cytoarchitecture. bioRxiv preprint 2020.08.06.219444 (2020).

\bibitem{lee_pseudolabel}
Lee, D.H.: Pseudo-Label: The Simple and Efficient Semi-Supervised Learning Method for Deep Neural Networks. In: ICML 2013 Workshop: Challenges in Representation Learning (WREPL). (2013).

\bibitem{mazzamuto_automatic}
Mazzamuto, G. et al.: Automatic Segmentation of Neurons in 3D Samples of Human Brain Cortex. In: Applications of Evolutionary Computation. pp. 78–85 (2018).

\end{thebibliography}
%

\end{document}